\documentclass[10pt,a4paper,onecolumn]{article}
\usepackage{marginnote}
\usepackage{graphicx}
\usepackage{xcolor}
\usepackage{authblk,etoolbox}
\usepackage{titlesec}
\usepackage{calc}
\usepackage{tikz}
\usepackage{hyperref}
\hypersetup{colorlinks,breaklinks,
            urlcolor=[rgb]{0.0, 0.5, 1.0},
            linkcolor=[rgb]{0.0, 0.5, 1.0}}
\usepackage{caption}
\usepackage{tcolorbox}
\usepackage{amssymb,amsmath}
\usepackage{ifxetex,ifluatex}
\usepackage{seqsplit}
\usepackage{fixltx2e} % provides \textsubscript
\usepackage[
  backend=biber,
]{biblatex}
\bibliography{paper.bib}

\usepackage[top=3.5cm, bottom=3cm, right=1.5cm, left=1.0cm,
            headheight=2.2cm, reversemp, includemp, marginparwidth=4.5cm]{geometry}

\titleformat{\section}
  {\normalfont\sffamily\Large\bfseries}
  {}{0pt}{}
\titleformat{\subsection}
  {\normalfont\sffamily\large\bfseries}
  {}{0pt}{}
\titleformat{\subsubsection}
  {\normalfont\sffamily\bfseries}
  {}{0pt}{}
\titleformat*{\paragraph}
  {\sffamily\normalsize}

\usepackage{fancyhdr}
\makeatletter
\let\ps@plain\ps@fancy
\fancyheadoffset[L]{4.5cm}
\fancyfootoffset[L]{4.5cm}

\definecolor{linky}{rgb}{0.0, 0.5, 1.0}

\newtcolorbox{repobox}
   {colback=red, colframe=red!75!black,
     boxrule=0.5pt, arc=2pt, left=6pt, right=6pt, top=3pt, bottom=3pt}

\newcommand{\ExternalLink}{%
   \tikz[x=1.2ex, y=1.2ex, baseline=-0.05ex]{%
       \begin{scope}[x=1ex, y=1ex]
           \clip (-0.1,-0.1)
               --++ (-0, 1.2)
               --++ (0.6, 0)
               --++ (0, -0.6)
               --++ (0.6, 0)
               --++ (0, -1);
           \path[draw,
               line width = 0.5,
               rounded corners=0.5]
               (0,0) rectangle (1,1);
       \end{scope}
       \path[draw, line width = 0.5] (0.5, 0.5)
           -- (1, 1);
       \path[draw, line width = 0.5] (0.6, 1)
           -- (1, 1) -- (1, 0.6);
       }
   }

\patchcmd{\@maketitle}{center}{flushleft}{}{}
\patchcmd{\@maketitle}{center}{flushleft}{}{}
\patchcmd{\@maketitle}{\LARGE}{\LARGE\sffamily}{}{}
\def\maketitle{{%
  
  \AB@maketitle}}
\makeatletter
\renewcommand\AB@affilsepx{ \protect\Affilfont}
\renewcommand\AB@affilnote[1]{{\bfseries #1}\hspace{3pt}}
\makeatother

\renewcommand\Affilfont{\sffamily\small\mdseries}
\ifnum 0\ifxetex 1\fi\ifluatex 1\fi=0 % if pdftex
  \usepackage[T1]{fontenc}
  \usepackage[utf8]{inputenc}

\else % if luatex or xelatex
  \ifxetex
    \usepackage{mathspec}
  \else
    \usepackage{fontspec}
  \fi
  \defaultfontfeatures{Ligatures=TeX,Scale=MatchLowercase}

\fi
\IfFileExists{upquote.sty}{\usepackage{upquote}}{}
\IfFileExists{microtype.sty}{%
\usepackage{microtype}
\UseMicrotypeSet[protrusion]{basicmath} % disable protrusion for tt fonts
}{}

\usepackage{hyperref}
\hypersetup{unicode=true,
            pdftitle={kima: Exoplanet detection in radial velocities},
            pdfborder={0 0 0},
            breaklinks=true}
\usepackage{graphicx,grffile}
\makeatletter
\def\maxwidth{\ifdim\Gin@nat@width>\linewidth\linewidth\else\Gin@nat@width\fi}
\def\maxheight{\ifdim\Gin@nat@height>\textheight\textheight\else\Gin@nat@height\fi}
\makeatother
\setkeys{Gin}{width=\maxwidth,height=\maxheight,keepaspectratio}
\IfFileExists{parskip.sty}{%
\usepackage{parskip}
}{% else
\setlength{\parindent}{0pt}
\setlength{\parskip}{6pt plus 2pt minus 1pt}
}
\ifx\paragraph\undefined\else
\let\oldparagraph\paragraph
\renewcommand{\paragraph}[1]{\oldparagraph{#1}\mbox{}}
\fi
\ifx\subparagraph\undefined\else
\let\oldsubparagraph\subparagraph
\renewcommand{\subparagraph}[1]{\oldsubparagraph{#1}\mbox{}}
\fi

\title{kima: Exoplanet detection in radial velocities}

        \author[1, 2]{J. P. Faria}
          \author[1, 2]{N. C. Santos}
          \author[1]{P. Figueira}
          \author[3]{B. J. Brewer}
    
      \affil[1]{Instituto de Astrofísica e Ciências do Espaço, Universidade do Porto,
CAUP, Rua das Estrelas, 4150-762, Porto, Portugal}
      \affil[2]{Departamento de Física e Astronomia, Faculdade de Ciências, Universidade
do Porto, Rua do Campo Alegre, 4169-007, Porto, Portugal}
      \affil[3]{Department of Statistics, The University of Auckland, Private Bag 92019,
Auckland 1142, New Zealand}
  \date{\vspace{-5ex}}

\begin{document}
\maketitle

\marginpar{
  %\hrule
  \sffamily\small

  {\bfseries DOI:} \href{https://doi.org/10.21105/joss.00487}{\color{linky}{10.21105/joss.00487}}

  \vspace{2mm}

  {\bfseries Software}
  \begin{itemize}
    \setlength\itemsep{0em}
    \item \href{https://github.com/openjournals/joss-reviews/issues/487}{\color{linky}{Review}} \ExternalLink
    \item \href{https://github.com/j-faria/kima}{\color{linky}{Repository}} \ExternalLink
    \item \href{dx.doi.org/10.6084/m9.figshare.6615350}{\color{linky}{Archive}} \ExternalLink
  \end{itemize}

  \vspace{2mm}

  {\bfseries Submitted:} 04 December 2017\\
  {\bfseries Published:} 19 June 2018

  \vspace{2mm}
  {\bfseries Licence}\\
  Authors of papers retain copyright and release the work 
  under a Creative Commons Attribution 4.0 International License 
  (\href{http://creativecommons.org/licenses/by/4.0/}{\color{linky}{CC-BY}}).
}

\section{Summary}\label{summary}

The radial-velocity (RV) method is one of the most successful in the
detection of exoplanets. An orbiting planet induces a gravitational pull
on its host star, which is observed as a periodic variation of the
velocity of the star in the direction of the line of sight. By measuring
the associated wavelength shifts in stellar spectra, a RV timeseries is
constructed. These data provide information about the presence of (one
or more) planets and allow for the planet mass(es) and several orbital
parameters to be determined (e.g. Fischer et al. 2016).

One of the main barriers to the detection of Earth-like planets with RVs
is the intrinsic variations of the star, which can easily mimic or hide
true RV signals of planets. Gaussian processes (GP) are now seen as a
promising tool to model the correlated noise that arises from
stellar-induced RV variations. (e.g. Haywood et al. 2014).

\textbf{kima} is a package for the detection and characterization of
exoplanets using RV data. It fits a sum of Keplerian curves to a
timeseries of RV measurements, using the Diffusive Nested Sampling
algorithm (Brewer, Pártay, and Csányi 2011) to sample from the posterior
distribution of the model parameters. This algorithm can sample the
multimodal and correlated posteriors that often arise in this problem
(e.g. Brewer and Donovan 2015).

Unlike similar open-source packages, \textbf{kima} calculates the fully
marginalized likelihood, or \emph{evidence}, both for a model with a
fixed number \(N_p\) of Keplerian signals, or after marginalising over
\(N_p\). For this latter task, \(N_p\) itself is a free parameter and we
sample from its posterior distribution using the trans-dimensional
method proposed by Brewer (2014). Because \textbf{kima} uses the
Diffusive Nested Sampling algorithm, the evidence values are still
accurate when the likelihood function contains phase changes which would
make other algorithms (such as thermodynamic integration) unreliable
(Skilling 2006).

Moreover, \textbf{kima} can use a GP with a quasi-periodic kernel as a
noise model, to deal with activity-induced signals. The hyperparameters
of the GP are inferred together with the orbital parameters. Priors for
each of the parameters can be easily set by the user, with a broad
choice of standard probability distributions already implemented.

The code is written in C\texttt{++}, but also includes a helper Python
package, \texttt{pykima}, which facilitates the analysis of the results.
It depends on the \texttt{DNest4} and the \texttt{Eigen} packages, which
are included as submodules in the repository. Other (Python)
dependencies are the \texttt{numpy}, \texttt{scipy},
\texttt{matplotlib}, and \texttt{corner} packages. Documentation can be
found in the main repository, that also contains a set of examples of
how use \textbf{kima}, serving as the package's test suite.

Initial versions of this package were used in the analysis of HARPS RV
data of the active planet-host CoRoT-7 (Faria et al. 2016), in which the
orbital parameters of the two exoplanets CoRoT-7b and CoRoT-7c, as well
as the rotation period of the star and the typical lifetime of active
regions, were inferred from RV observations alone.

\subsection{Acknowledgements}\label{acknowledgements}

This work was supported by Fundação para a Ciência e a Tecnologia (FCT)
through national funds and FEDER through COMPETE2020, by the grants
UID/FIS/04434/2013 \& POCI-01-0145-FEDER-007672 and
PTDC/FIS-AST/1526/2014 \& POCI-01-0145-FEDER-016886. J.P.F. acknowledges
support from the fellowship SFRH/BD/93848/2013 funded by FCT (Portugal)
and POPH/FSE (EC). N.C.S. and P.F. also acknowledge support from FCT
through Investigador FCT contracts IF/00169/2012/CP0150/CT0002 and
IF/01037/2013/CP1191/CT0001, respectively. B.J.B. acknowledges support
from the Marsden Fund of the Royal Society of New Zealand.

\begin{figure}
\centering
\includegraphics{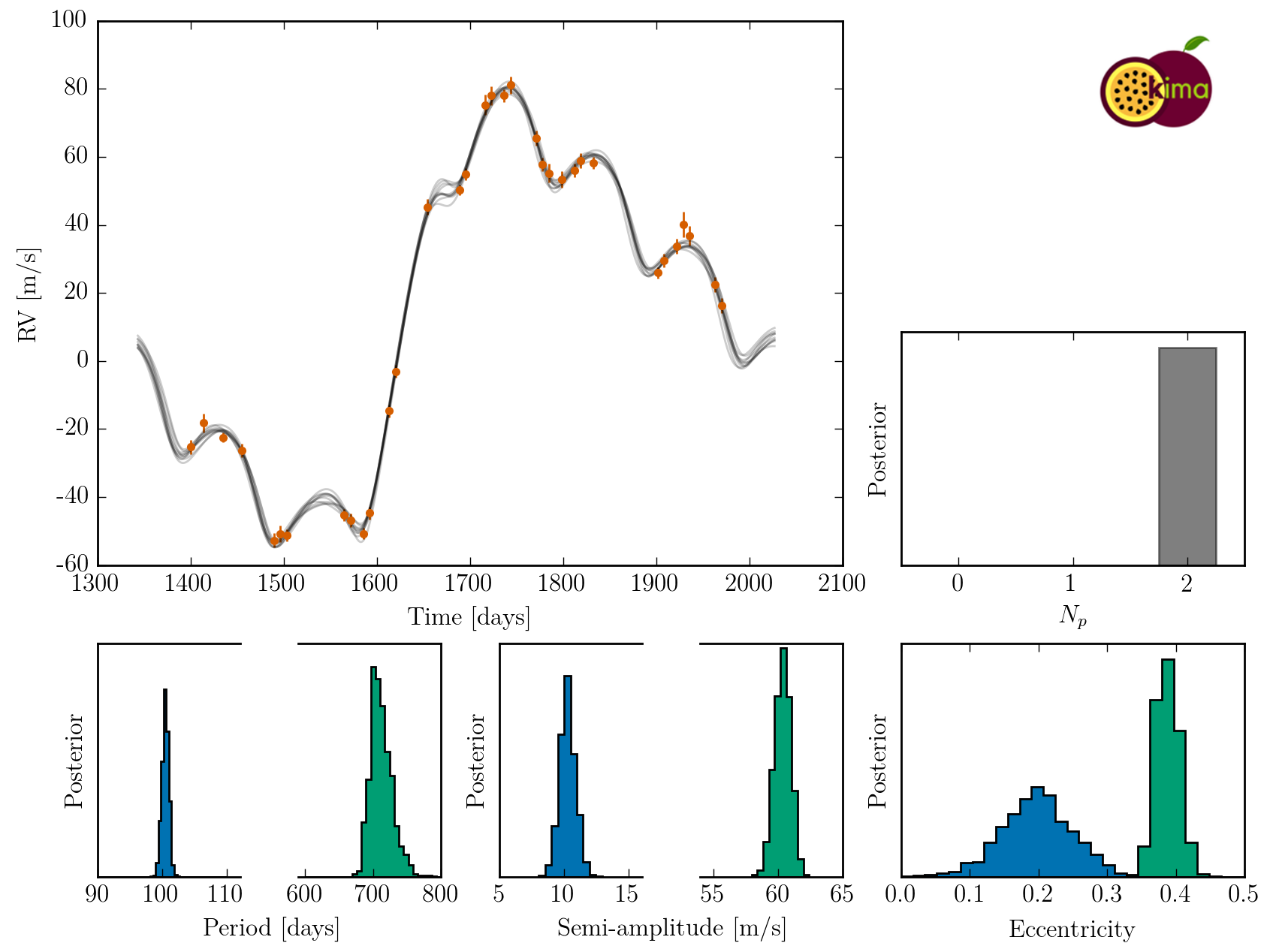}
\caption{Results from a typical analysis with \textbf{kima}. The top
panel shows a simulated RV dataset with two injected ``planets''. The
black curves represent ten samples from the posterior predictive
distribution. On the top right, the posterior distribution for the
number of planets \(N_p\) clearly favours the detection of the two
planets (this parameter had a uniform prior between 0 and 2). The panels
in the bottom row show the posterior distributions for the orbital
periods, the semi-amplitudes and the eccentricities of the two signals.}
\end{figure}

\section{References}\label{references}

\hypertarget{refs}{}
\hypertarget{ref-Brewer2014}{}
Brewer, B. J. 2014. ``Inference for Trans-Dimensional Bayesian Models
with Diffusive Nested Sampling.'' \emph{arXiv:1411.3921}, November.
\url{http://arxiv.org/abs/1411.3921}.

\hypertarget{ref-Brewer2015}{}
Brewer, B. J., and C. P. Donovan. 2015. ``Fast Bayesian Inference for
Exoplanet Discovery in Radial Velocity Data.'' \emph{Monthly Notices of
the Royal Astronomical Society} 448 (4): 3206--14.
doi:\href{https://doi.org/10.1093/mnras/stv199}{10.1093/mnras/stv199}.

\hypertarget{ref-Brewer2011}{}
Brewer, B. J., L. B. Pártay, and G. Csányi. 2011. ``Diffusive Nested
Sampling.'' \emph{Statistics and Computing} 21 (4): 649--56.
doi:\href{https://doi.org/10.1007/s11222-010-9198-8}{10.1007/s11222-010-9198-8}.

\hypertarget{ref-Faria2016}{}
Faria, J. P., R. D. Haywood, B. J. Brewer, P. Figueira, M. Oshagh, A.
Santerne, and N. C. Santos. 2016. ``Uncovering the Planets and Stellar
Activity of CoRoT-7 Using Only Radial Velocities.'' \emph{Astronomy \&
Astrophysics} 588 (April): A31.
doi:\href{https://doi.org/10.1051/0004-6361/201527899}{10.1051/0004-6361/201527899}.

\hypertarget{ref-Fischer2016}{}
Fischer, D. A., G. Anglada-Escudé, P. Arriagada, R. V. Baluev, J. L.
Bean, F. Bouchy, L. A. Buchhave, et al. 2016. ``State of the Field:
Extreme Precision Radial Velocities.'' \emph{Publications of the
Astronomical Society of the Pacific} 128 (964): 066001.
doi:\href{https://doi.org/10.1088/1538-3873/128/964/066001}{10.1088/1538-3873/128/964/066001}.

\hypertarget{ref-Haywood2014}{}
Haywood, R. D., A. Collier Cameron, D. Queloz, S. C. C. Barros, M.
Deleuil, R. Fares, M. Gillon, et al. 2014. ``Planets and Stellar
Activity: Hide and Seek in the CoRoT-7 System.'' \emph{Monthly Notices
of the Royal Astronomical Society} 443 (September): 2517--31.
doi:\href{https://doi.org/10.1093/mnras/stu1320}{10.1093/mnras/stu1320}.

\hypertarget{ref-Skilling2006}{}
Skilling, John. 2006. ``Nested Sampling for General Bayesian
Computation.'' \emph{Bayesian Analysis} 1 (4). International Society for
Bayesian Analysis: 833--59.
doi:\href{https://doi.org/10.1214/06-BA127}{10.1214/06-BA127}.

\end{document}